\documentclass[lettersize,journal]{IEEEtran}
\usepackage{amsmath,amsfonts}
\usepackage{algorithmic}
\usepackage{algorithm}
\usepackage{array}
\usepackage{caption}
\usepackage{subfig}
\usepackage{textcomp}
\usepackage{stfloats}
\usepackage{url}
\usepackage{verbatim}
\usepackage{graphicx}
\usepackage{xcolor}
\usepackage{cite}
\usepackage{makecell,colortbl}
\definecolor{LightCyan}{rgb}{0.88,1,1}
\begin{document}

\title{Exploring the 6G Potentials: Immersive, Hyper Reliable, and Low-Latency Communication}
\author{Afsoon~Alidadi~Shamsabadi,~\IEEEmembership{Senior Member,~IEEE,}
Animesh~Yadav,~\IEEEmembership{Senior Member,~IEEE,}
Yasser~Gadallah,~\IEEEmembership{Senior Member,~IEEE,} and 
Halim~Yanikomeroglu,~\IEEEmembership{Fellow,~IEEE,}}
\maketitle

\begin{abstract}
The transition towards the sixth-generation~(6G) wireless telecommunications networks introduces significant challenges for researchers and industry stakeholders. The 6G technology aims to enhance existing usage scenarios through supporting innovative applications that require stringent key performance indicators~(KPIs). In some critical use cases of 6G, multiple KPIs, including immersive throughput, with an envisioned peak data rate of $\mathbf{1}$ Tbps, hyper-reliability, in the range of $\mathbf{10^{-5}}$ to $\mathbf{10^{-7}}$, and hyper low-latency, between $\mathbf{0.1}$ and $\mathbf{1}$ ms, must be achieved simultaneously to deliver the expected service experience. However, this is challenging due to the conflicting nature of these KPIs. This article proposes a new service class of 6G as immersive, hyper reliable, and low-latency communication~(IHRLLC), and introduces a potential network architecture to achieve the associated KPIs. Specifically, enhanced technologies, such as ultra-massive multiple-input multiple-output~(umMIMO)-aided \mbox{terahertz~(THz)} \mbox{communications}, \mbox{reconfigurable intelligent surfaces~(RIS)}, and \mbox{non-terrestrial networks~(NTN),} are viewed as the key enablers for achieving \mbox{immersive} data rates and hyper reliability. Given the computational complexity involved in employing these technologies, we propose mathematical and \mbox{computational} enabling technologies, such as learn-to-optimize~(L2O), generative-AI~(GenAI), quantum computing, and network digital twin~(NDT), to complement the proposed architecture and optimize the latency.
\end{abstract}

\section{Introduction}
Recently, the fifth-generation (5G) networks have become operational in many countries worldwide, while researchers are working on studies related to the sixth-generation (6G) wireless networks. These studies are inspired by potential use cases, requirements, and unresolved challenges. The previous generations of wireless networks have already provided significant enhancements that enabled a vast range of novel applications compared to those supported by legacy networks. Nevertheless, according to the International Telecommunication Union's (ITU) vision for international mobile telecommunications 2030 (IMT-2030), 6G usage scenarios are envisioned as enhanced versions of existing 5G scenarios, alongside entirely new use cases that support the deployment of innovative applications \cite{IMT-2030}. Additionally, 6G wireless networks are expected to cater to future use cases and applications that require the concurrent fulfillment of high mobility, immersive data rates, hyper-reliability\footnote{Reliability, as defined by ITU, is the ability to transmit a predefined amount of data successfully within a specified time and probability.}, and low-latency. This necessitates the introduction of significantly more advanced network architectures and enabling technologies. We refer to this type of service class as immersive, hyper reliable, and low-latency communication (IHRLLC), which is also the focus of this article.

The current 5G wireless network already includes ultra-reliable and low latency communications (URLLC) as one of its service classes. The URLLC class targets a wide range of use cases, including industry automation, intelligent transportation, health care, and intelligent societies.
Several enabling schemes and technologies have been proposed to meet the stringent requirements of URLLC in terms of \mbox{reliability} and latency~\cite{new_uRLLC}. These include grant-free access, spatial diversity, short packet transmission, slicing, edge caching and computing, etc.
However, the critical use cases and applications that are envisioned for the 6G wireless networks demand unprecedented data rates combined with hyper-low latency and reliability simultaneously. In this context, URLLC alone might not be sufficient to achieve the requirements set for IHRLLC. Furthermore, it is essential to note that the IHRLLC requirements conflict with each other, posing significant challenges. Therefore, novel network architecture designs, along with innovative schemes and technologies, must be explored and investigated.

Recently, a few works have started investigating the role of URLLC in addressing use cases and applications that require high connectivity, mobility, and high data rates in 6G wireless networks~\cite{Chaccour-Soorki-Saad-Bennis-opovski-IEEE-IOTJ-2022,
Liu-Deng-Nallanathan-Yuan-IEEE-WCM-2023, Yuan-Zou-Cui-Li-Mu-Han-IEEE-WCM-2023}. In \cite{Chaccour-Soorki-Saad-Bennis-opovski-IEEE-IOTJ-2022}, the potential of the terahertz (THz) frequency band is explored for offering  high-rate and highly-reliable low-latency
communications in wireless virtual reality (VR) applications. 
In \cite{Liu-Deng-Nallanathan-Yuan-IEEE-WCM-2023}, three connectivity scenarios requiring high data rates, namely ubiquitous, deep, and holographic connectivity, are proposed, leveraging URLLC.
In \cite{Yuan-Zou-Cui-Li-Mu-Han-IEEE-WCM-2023}, a high mobility communication scenario is considered, and the impact of the Doppler spread on URLLC for vehicular communication is explored.
The aforementioned works explore individual use cases that require high data rates while adhering to the URLLC constraints. In this article, we present several critical use cases for 6G wireless networks under the proposed IHRLLC service class. Additionally, we discuss potential enabling technologies and propose a unified approach to integrate these technologies to support the IHRLLC services. The main contributions of this paper can be summarized as follows:
\begin{itemize}
    \item We propose a new type of service class for 6G, namely the IHRLLC class, which concurrently supports an \mbox{immersive} data rate, hyper reliability, and low-latency communication. Several important use cases for this type of service are discussed, along with their specific \mbox{requirements}.  
    \item We discuss a set of enabling technologies to support the IHRLLC class. Additionally, a network architecture is proposed which combines these technologies to achieve the goal of IHRLLC. Challenges and open issues pertaining to the proposed architecture are also discussed.
    \item We present a numerical simulation-based experiment to validate the high throughput via ultra-massive multiple-input multiple-output (umMIMO)-aided THz \mbox{communication}.
\end{itemize}
\section{Requirements, Use Cases, and Challenges}\label{Use Case}
As 6G wireless networks are expected to encompass a broad spectrum of novel use cases, applications, and services, more than one KPI must be met simultaneously to achieve acceptable levels of quality of experience (QoE). Considering the end-user experience, QoE can be defined as ``the \mbox{overall} acceptability of an application or service, as perceived subjectively by the end-user" \cite{6GQoE}. In the following section, we discuss a novel service class for 6G which we term the IHRLLC class, and discuss some key use cases and associated challenges. Specifically, we focus on use cases that require the simultaneous satisfaction of three KPIs, namely, immersive data rates, hyper reliability, and low-latency. 
\subsection{Immersive Communication}
 Immersive communication (IC) will be one of the key service classes of the 6G wireless networks. It will introduce interactive services, such as extended reality (XR), holographic communication, cloud gaming, high-definition video (4K and 8K) streaming, and immersive event experiences over cellular networks. These services demand a large volume of data and require immersive throughput values. For instance, a live VR concert, streaming to thousands of participants simultaneously, requires a wireless network capable of handling substantial data rates without compromising the user's QoE. According to the ITU IMT-2030 vision, the envisioned throughput for IC peaks at $1$ Tbps, with the user-experienced data rate of at least $300$ Mbps\cite{IMT-2030}, which far exceeds the capabilities of IMT-2020.
\subsection{Hyper Reliable and Low-Latency Communication (HRLLC)}
The HRLLC class, an extension of URLLC in 5G, will be another core service class in 6G.
Latency in wireless networks is influenced by various components of the architecture. Key factors include network management strategies in the core network, signal processing schemes, such as modulation and coding in the end-to-end (E2E) parts of the network, and the propagation delay of the signal within the network. Achieving hyper low-latency values of $0.1$ to $1$ ms, as specified by IMT-2030, without ensuring a hyper-reliable network, renders 6G use cases infeasible. Specifically, for applications, such as autonomous vehicles and industrial automation, the network must simultaneously meet the reliability standard of ${10^{-5}}$ to ${10^{-7}}$ while achieving hyper low-latency, thus enabling the HRLLC. 
\subsection{Immersive, Hyper Reliable, Low-Latency Communication (IHRLLC)}
The two aforementioned service classes have been studied separately, each with its own associated KPIs. Nevertheless, a wide range of use cases requires simultaneous fulfillment of the three KPIs (i.e., immersive data rate, hyper reliability, and low-latency) to achieve an acceptable level of QoE. Hence, we propose IHRLLC as a new service class for 6G wireless networks. The IHRLLC service class must ensure that the three KPIs are consistently met, with zero tolerance for service interruptions. Below are some of the key applications under this service class.

\subsubsection{Mobile tele-surgery} Mobile tele-surgery in an ambulance racing through the city with a patient needing an urgent surgical operation is a representative use case for IHRLLC. 
This situation requires not only IC for transmitting substantial data volumes but also absolute reliability and hyper low latency to ensure the safety and efficacy of the medical procedure.

\subsubsection{Multisensory XR} XR refers to various types of reality, including augmented reality, mixed reality, and VR.
Multisensory XR applications have unique requirements, including bounded low latency and high uplink and downlink bandwidth.
The 3rd Generation Partnership Project (3GPP) Release 18 promises to significantly improve support for higher data rates and capacity while also introducing power and latency reduction techniques to enable more immersive XR experiences.

\subsubsection{High-frequency trading} High-frequency trading (HFT) is a trading method that utilizes powerful computer programs to execute a large number of orders in a very short period. This approach aims to avoid the high risks associated with the fast-changing trading prices.
HFT requires networks that are capable of meeting the required set of KPIs with a high degree of certainty.

\subsubsection{Cloud gaming} In the gaming industry, cloud gaming refers to use cases where most computations related to single-player or multiplayer gaming are offloaded from users' hardware devices to edge or remote servers. Essentially, games are hosted on remote servers, with users' devices serving primarily as peripherals for interacting with the game. In cloud gaming, the user's QoE depends on the network's capability to deliver IHRLLC service class performance.

\subsubsection{Advance air mobility} Advance air mobility (AAM)
enables customers to access on-demand air transportation through multimodal transport networks. One of the key use cases of AAM is urban air mobility (UAM), which allows vehicles to either be autonomously controlled or remotely piloted while operating beyond the visual line-of-sight (LoS) from the ground. 
The UAM optimal performance requires real-time immersive data transmission with hyper reliability and low-latency to a ground pilot for remote operation.

Pursuing a combined set of conflicting KPIs leads to inherent difficulties, as most existing wireless networks usually focus on achieving a specific KPI at the expense of others. Balancing these KPIs in 6G wireless networks to meet the diverse requirements of IHRLLC remains a significant challenge, and addressing them necessitates innovative solutions. In the following section, we overview a set of potential enabling technologies and describe how properly aggregating these technologies overcomes the challenges that arise in simultaneously achieving conflicting KPIs.
\begin{figure}[t]
    \centering
    \captionsetup{justification=justified}
    \includegraphics[width=\columnwidth]{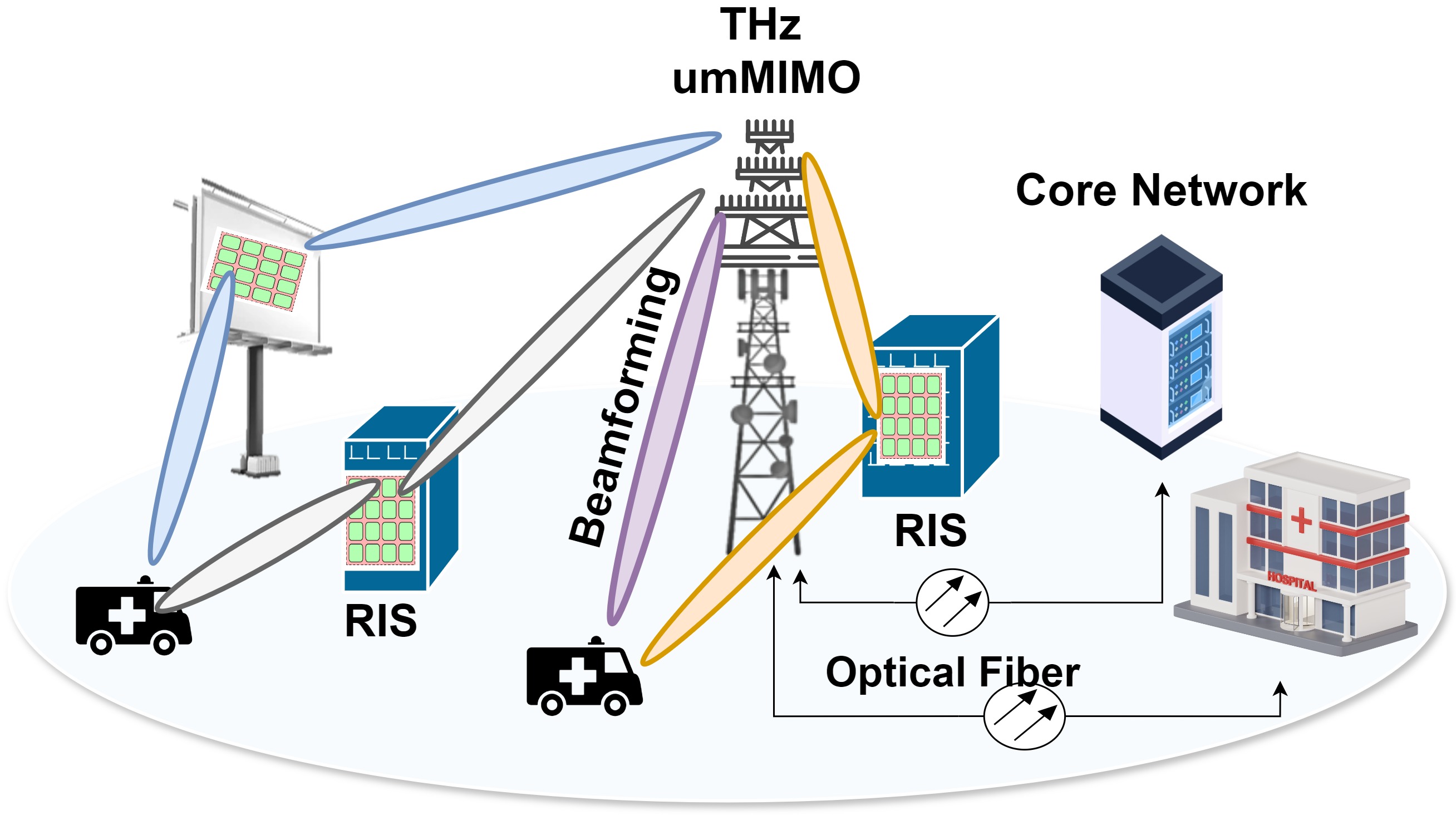}
    \caption{\small umMIMO and RIS-aided THz communication for 6G mobile tele-surgery application.}
    \label{fig:umMIMO and RIS and THz}
\end{figure}
\section{Envisioned Enabling Technologies for IHRLLC}\label{Technologies}
Several solutions and technologies are proposed to improve the throughput, latency, and reliability of wireless networks. These technologies span different parts of the wireless networks to ensure that the overall requirements are met. However, techniques used to achieve one specific KPI may interfere with achieving other KPIs. For instance, utilizing higher frequency bands significantly improves data rates but limits the coverage area due to severe shadowing, high path loss, and molecular absorption, ultimately affecting reliability. Similarly, techniques like edge computing, which prioritize low latency, may struggle to scale effectively in scenarios requiring hyper-reliable communication. In this section, we provide an overview of our envisioned IHRLLC enabling technologies, highlighting their advantages and challenges. To address the trade-offs between different KPIs, we propose leveraging complementary technologies that work together to mitigate these challenges. Additionally, we discuss how integrating diverse technologies across various network layers can facilitate the simultaneous achievement of multiple KPIs.
\subsection{umMIMO and RIS-aided THz Communication}
Migrating to higher frequency bands has been the solution to access wider bandwidth, thereby enabling higher data rates for users. The millimeter wave (mmWave) spectrum, which encompasses frequencies above $24$ GHz, was utilized in the 5G wireless networks to provide access to a considerable bandwidth compared to earlier mobile network generations. However, the immense capacity demands of 6G applications necessitate access to even broader bandwidths, achievable only through THz communication. The THz spectrum spans frequencies from $0.1$ to $3$ THz. Extremely high available bandwidth at this frequency range makes it a promising technology to achieve immersive communication in 6G. Nevertheless, operating in the THz frequency band exposes new challenges to the wireless networks \cite{THzSurvey}. The most significant challenges include higher free space path loss (FSPL), molecular absorption, channel modeling, and the design of hardware components, such as radio frequency (RF) devices, modulators, and antennas. Therefore, innovative solutions are essential to address the aforementioned challenges and fully harness the potential of THz frequencies in 6G wireless networks. For instance, to overcome the short communication range and severe blockage issues associated with the THz band, advanced antenna array architectures, e.g., umMIMO and reconfigurable intelligent surfaces (RIS), can be integrated into radio access networks to improve the beam gain toward users and help maintain LoS connection between the base stations and users, respectively~(Fig.~\ref{fig:umMIMO and RIS and THz}). 
\begin{figure}[t]
    \centering
    \captionsetup{justification=justified}
    \includegraphics[width=0.9\columnwidth]{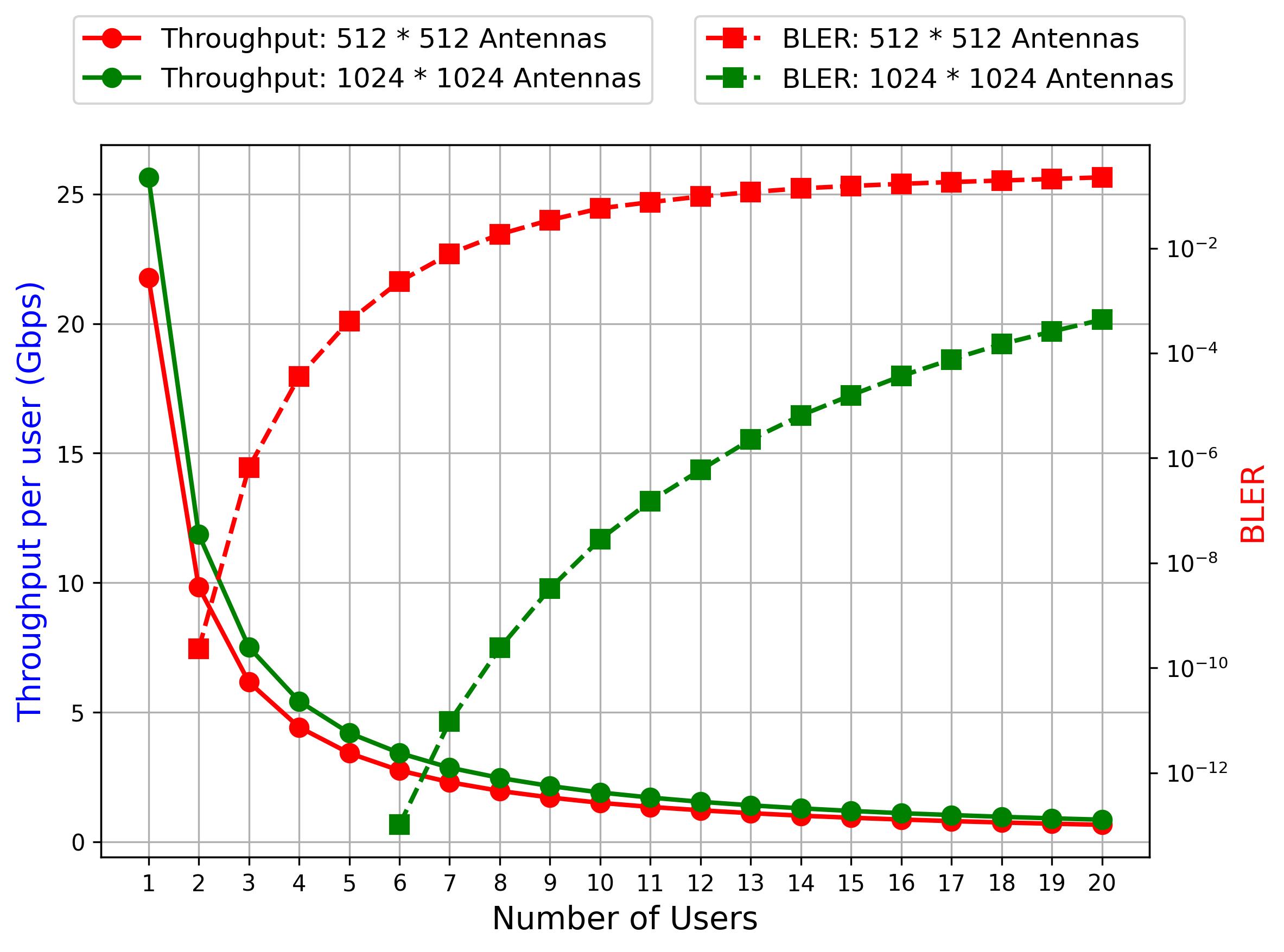}
    \caption{\small Per user throughput and BLER versus the number of users for umMIMO-enabled THz communication.}
    \label{fig:DataRateAndBER}
\end{figure}
\subsubsection{umMIMO beamforming}
Higher frequency bands, such as mmWave and THz, enable the use of umMIMO antenna designs that can accommodate a significantly greater number of antenna elements, compared to the traditional MIMO architectures. As a result of implementing the umMIMO technology and enhanced beamforming techniques, high-gain narrow antenna beams can be tailored to individual users. These beams help overcome the substantial FSPL and short-range communication in THz communications~\cite{umMIMO}. 
Fig.~\ref{fig:DataRateAndBER} plots the throughput and block error rate (BLER) performance under different number of antenna elements versus the number of users in the umMIMO-enabled THz communication system. The network consists of a base station and up to $20$ IHRLLC users in a cell of radius $500$ meters, operating at a center frequency of $850$ GHz with bandwidth of $2$ GHz. The network resources are equally and orthogonally divided among users to eliminate interference. It is observed that, with sufficient available resources, the throughput and BLER values will be within acceptable ranges, supporting immersive and reliable communication. The performance improves with higher-order umMIMO antennas; for instance, upgrading from a $512 \times 512$ antenna array to a $1024 \times 1024$ array yields slight throughput improvement. This moderate gain is influenced by the logarithmic nature of the relationship, but further enhancements can be expected with even higher-order umMIMO arrays. Nonetheless, achieving such high data rates and low BLER for a large number of users with limited resources highlights the necessity of advanced beamforming algorithms and strong error control coding techniques to optimize performance.

umMIMO beamforming in the THz frequency band presents several challenges, including the need for precise channel state information (CSI), accurate angle of departure of the transmitters, and the angle of arrival of receivers. Additionally, beamforming is highly sensitive to mobility, necessitating rapid and continuous beam realignment. Moreover, a large number of RF power amplifier chains are needed to perform effective beamforming, thus increasing the cost and power consumption. Alternatively, an energy-efficient holographic MIMO system appears to be a potential solution. 
\subsubsection{RIS for LoS connectivity and diversity}
The RIS technology presents a groundbreaking advancement in wireless communication, offering an evolutionary approach to signal propagation and control. RISs are artificial surfaces with numerous small, electronically controlled passive elements capable of altering the phase and amplitude of the incoming signal, thus enabling the redirection or reshaping of the signal path\cite{RIS_New}. Therefore, an RIS enables a favorable and adaptable wireless environment for signal propagation, which can enhance the performance of wireless networks, specifically those operating within the THz frequency band. RISs can be strategically positioned to create a scattering channel, which is desirable for enhancing the performance of umMIMO systems in THz communications. As a result, RIS systems can add a degree of freedom to umMIMO systems and improve the reliability of wireless networks. In addition, considering the IHRLLC service class use cases that require a seamless guaranteed level of KPIs, wireless networks require deploying more base stations to ensure reliable connectivity for users. However, this will bring higher energy consumption, higher deployment and maintenance costs, and a higher level of interference to deal with. RIS passive elements and their light weight make them capable of being deployed in many strategic spots in the city at a relatively low cost, which enables them to overcome the communication distance limitations of the THz frequency band by offering LoS links, thus paving the way for IHRLLC. 

\begin{figure}[t]
    \centering
    \captionsetup{justification=justified}
    \includegraphics[width=\columnwidth]{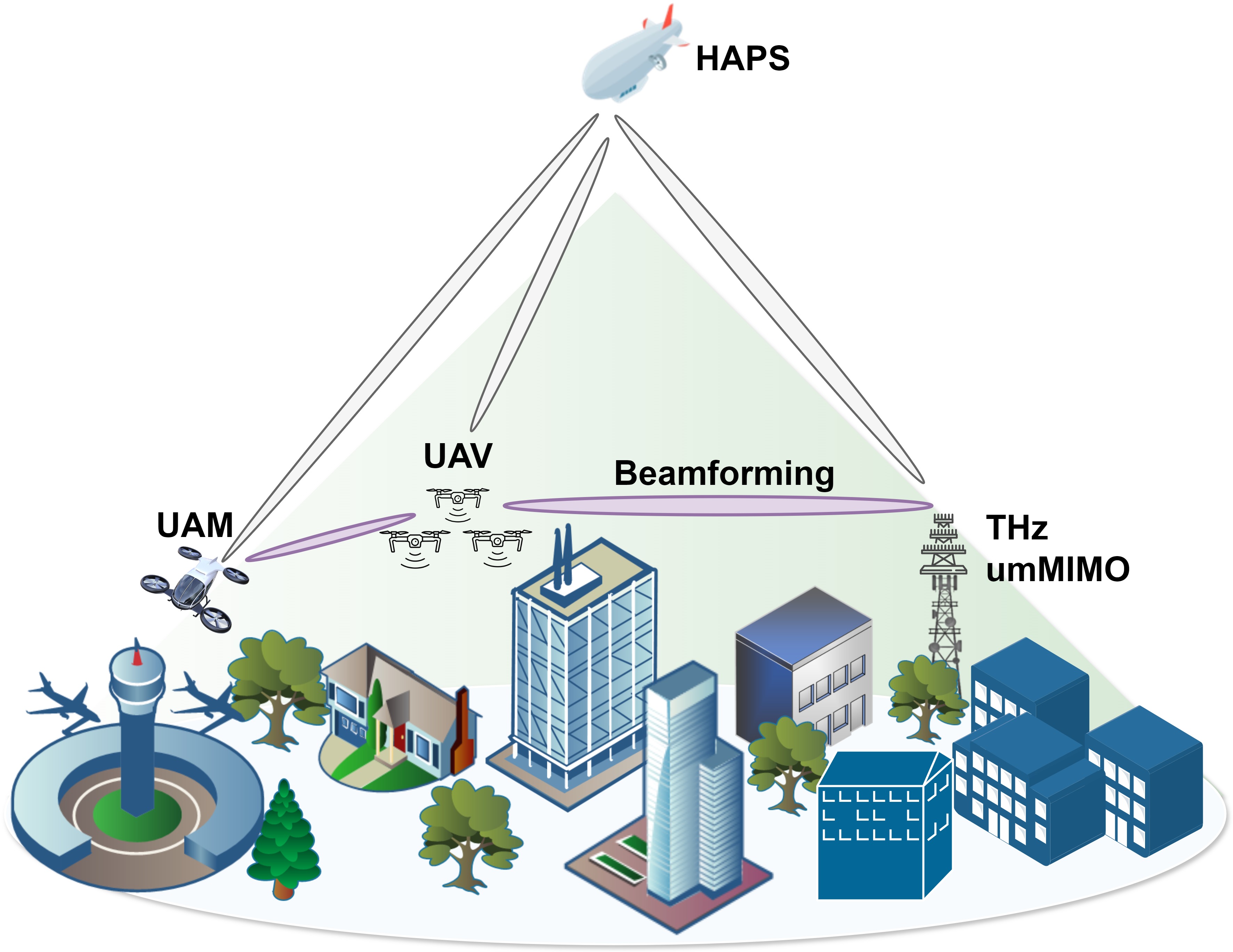}
    \caption{\small Role of HAPS and UAVs in enabling UAM use case in 6G.}
    \label{fig:NTN}
\end{figure}
\begin{figure*}[t]
    \centering
    \subfloat[\small A comparison between the architecture of classical optimization and optimization using L2O\cite{L2O}.\label{L2O}]{
        \includegraphics[width=0.89\linewidth]{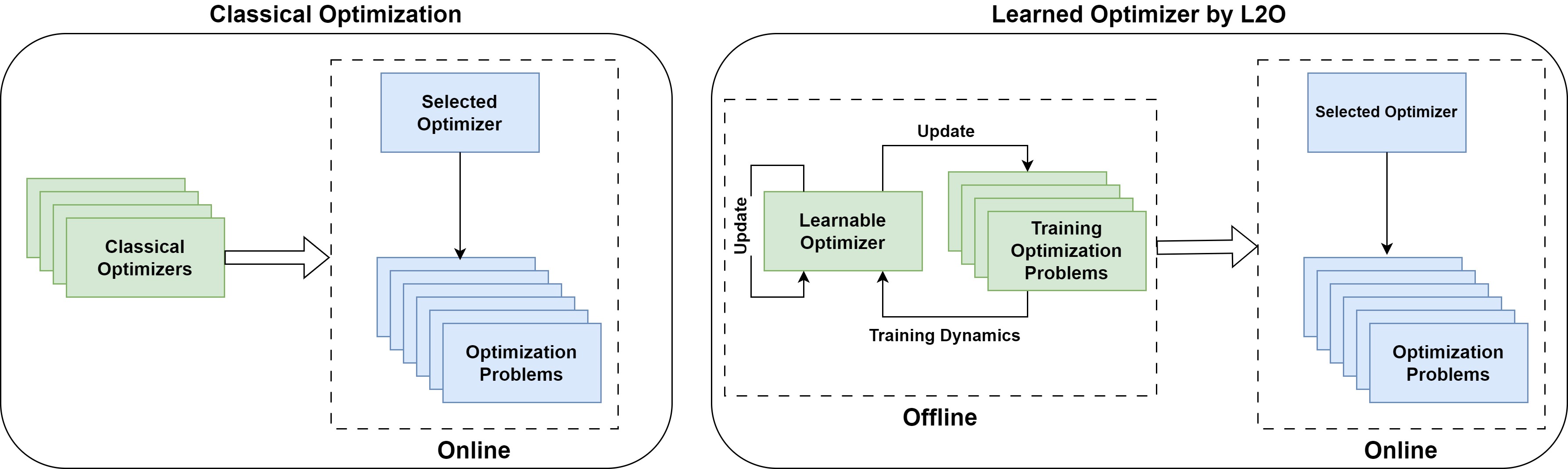}}\\
    \subfloat[\small Applications of GenAI in wireless \mbox{networks} physical layer\cite{GenAI}.\label{GenAI}]{
        \includegraphics[width=0.43\linewidth]{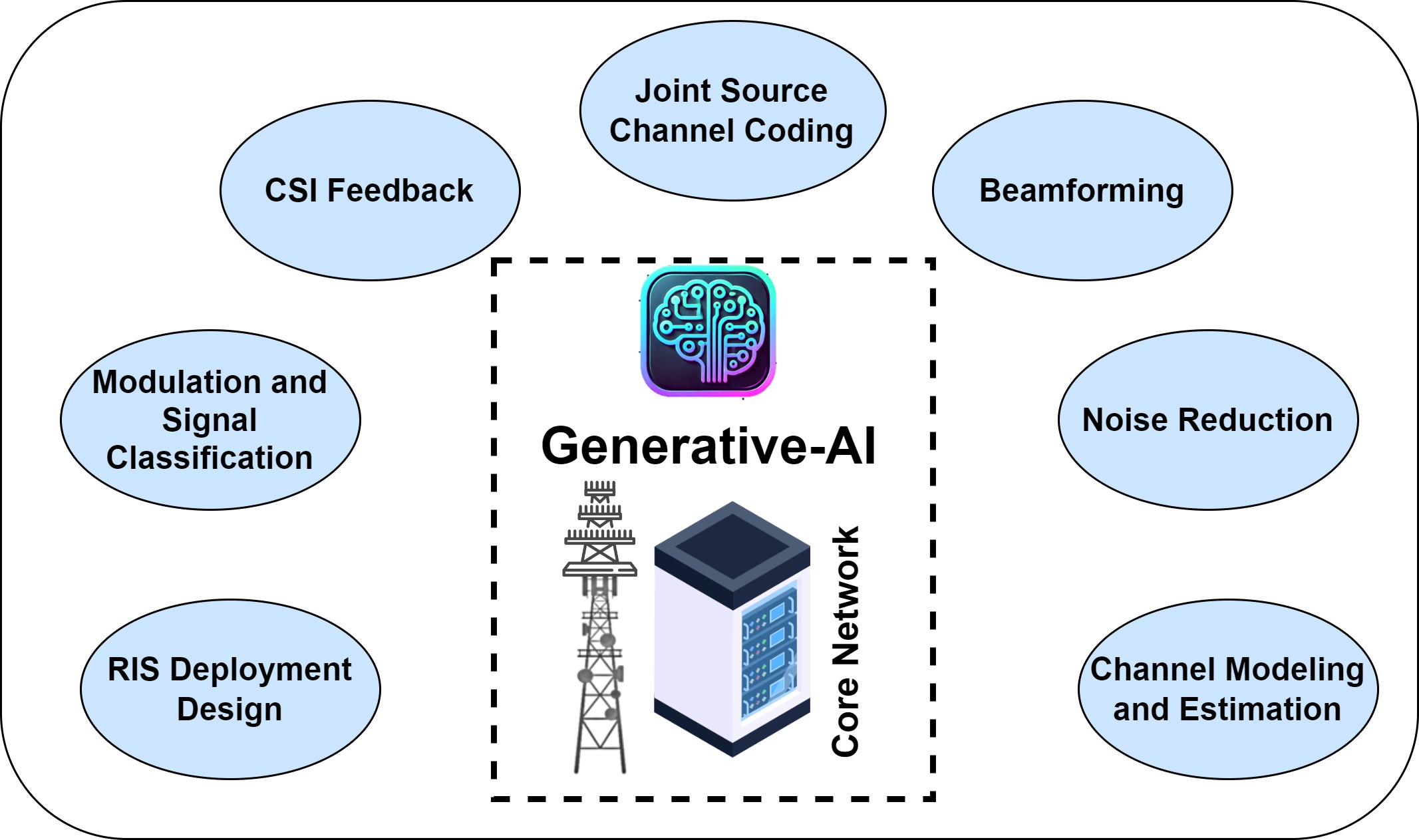}}
        \hspace{0.2cm}
    \subfloat[\small Quantum approximate optimization algorithm (QAOA) for combinatorial optimization.\label{Quantum Computing}]{
        \includegraphics[width=0.43\linewidth]{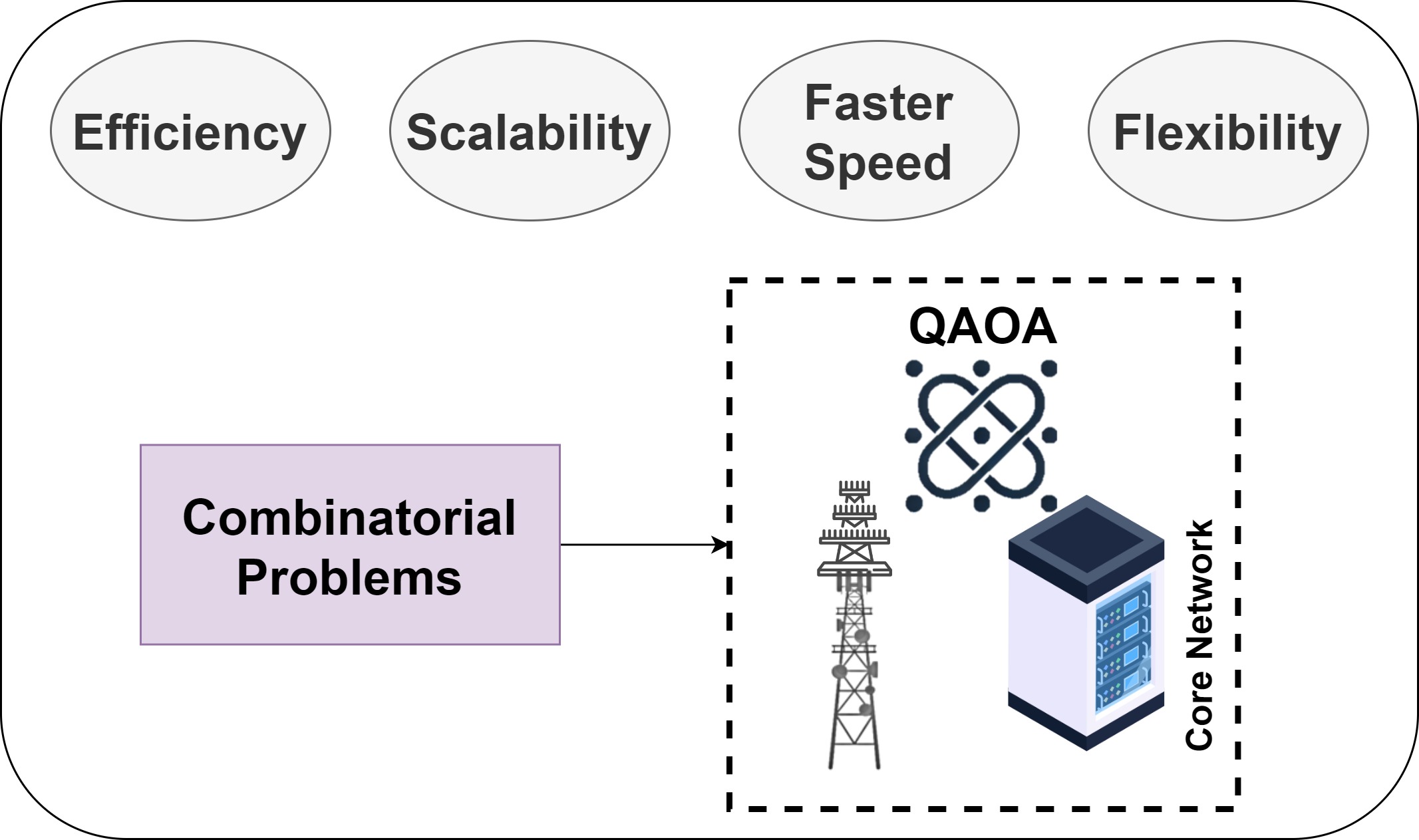}}
    \caption{\small L2O, GenAI, and Quantum Computing. The logos in the middle of (b) and (c) are generated using ChatGPT-4o.}
    \label{fig:L2O and GenAI, and Quantum}
\end{figure*}
\subsection{Multi-Connectivity through Non-Terrestrial Networks}
Non-terrestrial networks (NTN), which include space satellites and aerial platforms, such as high altitude platform stations (HAPS) and unmanned aerial vehicles (UAVs), present a promising architecture for 6G wireless networks. Traditionally, NTN have been considered a potential technology for coverage enhancement in rural and remote areas. However, given the ever-increasing capacity demands of future wireless networks and the need for a resilient and reliable network architecture, the new paradigm of employing the NTN is to integrate them into urban terrestrial wireless networks to guarantee an immersive and hyper reliable connection for the users~\cite{IM-RA2}. To this end, aerial platforms have the potential to be part of the proposed architecture for IHRLLC service class. As depicted in Fig.~\ref{fig:NTN}, in applications, such as UAM, multi-connectivity through aerial platforms, such as UAVs and HAPS, will be a potential solution in case of losing connection with either the main base station or the RIS. Moreover, HAPS, located at an altitude of about $20$ Km above the ground, can provide a one-way propagation delay of the order of microseconds, which is much less than that of the satellite networks. In addition, as depicted in Fig.~\ref{fig:NTN}, HAPS can play the role of backhauling for the UAVs by enabling airborne integrated access and backhauling (IAB). 

\subsection{AI Enablers: L2O and GenAI}
In the previous sections, various enabling technologies were discussed to facilitate IHRLLC. The implementation of such sophisticated technologies, as well as the proper design of the corresponding parameters, is a critical challenge. Considering the umMIMO antennas and RISs, with a large number of antenna elements along with the challenges of THz communication, designing the optimum parameters through traditional approaches incurs a high computational cost. To this end, AI algorithms can potentially produce revolutionary changes in future wireless networks and overcome computational challenges. In particular, AI can be integrated with wireless networks as a distinct network function or be embedded within existing functions, depending on network demands. In scenarios where AI needs to handle complex centralized tasks, AI can be deployed as a dedicated network function, responsible for high-level decision-making, and coordination with other functions to optimize the overall network performance. Alternatively, AI can be integrated as a module within existing network functions to enhance their operations. 
This modular integration enhances the intelligence and adaptability of current architectures without introducing a completely new function. In the following, we investigate promising novel AI enablers, including learn-to-optimize (L2O) and generative AI (GenAI) as methods to solve the aforementioned computational challenges in 6G wireless networks.

L2O leverages machine learning (ML) to improve current optimization algorithms or develop new optimization approaches \cite{L2O}. Conventional optimization techniques often rely on fundamental methods, such as gradient descent, conjugate gradient, and Newton steps, which are theoretically robust and straightforward. These methods are chosen based on the type of problem and are expected to provide solutions that meet their theoretical guarantees. However, L2O presents a different approach in which optimization methods are developed through training based on their performance with sample problems (Fig.~\ref{L2O}). These methods may not have a solid theoretical foundation but are designed to enhance performance through training, which usually occurs online. At the same time, this training aims to make the actual application of the methods more efficient. In situations involving complex problems, such as nonconvex optimization or inverse problems, a well-trained L2O method might deliver better results than traditional methods.

Although conventional AI/ML algorithms have already been utilized in the development of algorithms for wireless networks, GenAI has recently been considered a potential tool for future wireless networks. GenAI can be employed in wireless networks to overcome the challenges of the conventional AI algorithms~\cite{GenAI}. Specifically, GenAI models can be utilized in various physical layer aspects of the networks, such as RIS deployment design, joint source-channel coding, beamforming, etc., as illustrated in Fig.~\ref{GenAI}. In addition, GenAI models can generate data for training purposes and can be implemented near training locations. 
\vspace{-0.2cm}
\subsection{Quantum Computing}
Next-generation wireless networks will experience massive communications among users in various usage scenarios. This significantly complicates the decision-making process for determining the order of resource allocation. Additionally, these networks will leverage AI and cognitive technologies for self-healing, self-management, and decision making, requiring massive computing capabilities~\cite{QCN_IEEECM_2021}. These challenges can be framed as combinatorial problems which involve determining the service order for different users and the routes for traffic transfer to ensure that everyone receives the promised QoE.

Traditionally, combinatorial problems have been largely solved using heuristic methods that simplify the problem. However, these methods typically yield approximate answers that, in certain situations, may lead to inferior performance. On the other hand, optimization methods based on statistics and data, while more accurate, are energy-intensive and unsuitable for dynamic situations requiring quick responses. To address these limitations, hybrid algorithms, like the quantum approximate optimization algorithm (QAOA), that combine quantum and classical computing, offer a promising solution. These algorithms can be deployed in the core network at various scales, according to hardware capabilities, offering considerable performance improvements in solving combinatorial optimization problems \cite{Quantum_algo}. Hybrid algorithms ensure that demanding applications receive the necessary service quality, even during peak network usage, and also contribute to energy saving.
\vspace{-0.2cm}
\subsection{Network Digital Twin (NDT)}
Digital twin (DT) is an emerging technology that virtually represents a physical system. It accurately models complex systems by replicating real physical entities and functionalities of a physical system into the digital world in real-time. Combining the DT technology with AI/ML tools, real-time data can be leveraged to improve the operation and performance of the physical system through precise decision-making and efficient resource utilization. 
Recently, the network DT (NDT) has been proposed as a key enabler for efficient control and management of modern communication networks~\cite{NDT-IEEE-CM-2022}. NDT enables real-time monitoring of the status of edge and cloud servers. Additionally, integrating 3D map services with NDT allows for the determination of the location and health of network infrastructure elements, such as base stations and RISs. In this way, NDT significantly enhances network performance. However, it faces challenges, such as data asynchronization between the NDT and the real network. These challenges can be mitigated through the strategic application of AI/ML tools.
\begin{figure*}[t]
    \centering
    \captionsetup{justification=centering}
    \includegraphics[width=\linewidth]{Figures/Figure5.jpg}
    \caption{\small NDT-empowered network architecture for IHRLLC service class in 6G.}
    \label{fig:AllTech}
\end{figure*}
\vspace{-0.2cm}
\section{Proposed Network Architecture for IHRLLC}
We now propose a novel network architecture that integrates the aforementioned technologies to meet the requirements for the IHRLLC service class. The proposed NDT-empowered architecture,
designed for the mobile tele-surgery and the UAM use cases, is shown in Fig.~\ref{fig:AllTech}. This architecture benefits end users by enabling IHRLLC service while offering mobile operators and industries opportunities to enhance network architecture and develop new business models with improved QoE. In this architecture, the base stations are equipped with umMIMO antenna and communicate using the THz band to enable immersive communications. Given the short range communication nature of THz frequency band, numerous RISs are deployed at different locations throughout the city to reliably transmit data at high data rates. We use NDT as the underlying technology that helps the IHRLLC use cases to make more informed and swift decisions.
Obtaining the relative location information via NDT, and with the help of the umMIMO, the user continuously forms pencil beams toward multiple RISs in LoS with the user. The use of multiple RISs not only facilitates LoS communications, but also enhances diversity reception, which significantly improves reliability. Accordingly, each RIS is pre-tuned so that all reflected signals are directed to the nearest base station and subsequently to the core network. Furthermore, the proposed architecture provides multi-connectivity transmission through aerial platforms for the users. For instance, in the mobile tele-surgery use case, if the ambulance can not find an RIS in the range, it then establishes a connection through a dedicated UAV that is associated with this ambulance. For the UAM use case, UAVs and HAPS will be part of the proposed IHRLLC architecture, ensuring seamless connectivity for the user.

AI enablers (L2O and GenAI) and quantum computing play a crucial role in the proposed network architecture by reducing computational delays for efficient resource allocation, beamforming design, beam-tracking, and various management functions.
The L2O and GenAI models can be utilized at base stations, edge network (near the base station), and the core network, on different scales, to support the design of efficient algorithms. Accordingly, quantum computing resources are distributed across the edge nodes and core network components. In this way, tasks requiring heavy computation are delegated to the core network due to the higher processing capabilities. On the other hand, tasks requiring real-time decision making will be handled at the base stations using L2O and GenAI models. Task allocation is based on the status of all the servers, monitored via NDT. Additionally, in the proposed architecture, HAPS can serve as a powerful computing platform to offload the heavy non-real time computations from the base stations.
\section{Challenges and Open Issues}
Previous sections discuss the 6G use cases that require IHRLLC, and propose a combination of technologies and a unified network architecture to achieve IHRLLC with an acceptable QoE. However, designing an E2E wireless network capable of catering to IHRLLC use cases poses a significant challenge. Particularly, the developed algorithms in the proposed network architecture, namely, the resource allocation, beamforming design, user association, etc., should be capable of dealing with large-scale network parameters, related to the ultra massive number of antenna elements and RIS components. As a result, the scalability of the developed AI-driven algorithms is a critical challenge.
On the other hand, although the IHRLLC service class deals with the use cases that do not involve a large number of users, the algorithms should be scalable from the number of users perspective in order to ensure the expected QoE. In addition, efficient interference management approaches should be employed in the network to manage and mitigate the propagated interference among the users.
In addition to the challenges related to scalability, efficient implementation of the THz frequency band, combined with umMIMO and RIS technologies, necessitates the development of hardware devices (antenna elements and electronic/photonic devices) that can generate high frequency electromagnetic waves. This challenge is an open area of research among wireless researchers. 
In addition, the THz-specific air interface techniques and multiple access protocols need further exploration to ensure network reliability. Furthermore, human health and safety issues, associated with the THz technology, should be thoroughly analyzed and evaluated. To address this issue, regulatory policies should be considered while implementing the THz frequency band.

Although AI has been introduced as a game changer for future wireless networks, the specific characteristics of the wireless networks make the implementation of AI algorithms challenging. One of the key challenges is the randomness of the wireless network environments, inherited from the random channel models. This is considerably important in urban areas where wireless channels experience significant random small-scale fading.
This is in addition to the privacy and standardization concerns related to the implementation of AI algorithms. Finally, integrating a combination of all potential technologies into wireless networks necessitates interdisciplinary research among experts in wireless networks, AI, and quantum computing to ensure the efficiency and scalability of the developed algorithms. With a proper interdisciplinary coordination, the proposed IHRLLC service class and NDT-empowered network architecture have the potential to be part of the 6G standardization.
\section{Conclusion}\label{Conclusion}
In this paper, we proposed a new service class for 6G, referred to as immersive, hyper reliable, and low-latency communication (IHRLLC). The 6G use cases under the IHRLLC service class require achieving immersive throughput, with an envisioned peak data rate of $1$ Tbps, hyper-reliability, in the range of $10^{-5}$ to $10^{-7}$, and low-latency, with values between $0.1$ and $1$ ms, simultaneously. To achieve this, we proposed an NDT-empowered network architecture which incorporates potential technologies to meet the KPI requirements and overcome the challenges of simultaneously achieving multiple KPIs. To this end, THz communications, along with umMIMO and RISs, play an important role in enabling the IHRLLC class. In addition, novel computation and communication techniques, including L2O, GenAI, and quantum computing, were proposed to facilitate the design of the parameters of various parts of the architecture.
\bibliographystyle{IEEEtran}
\bibliography{Ref2}
\section*{Biography}
\vspace{-33pt}
\begin{IEEEbiographynophoto}{Afsoon Alidadi Shamsabadi}
    [SM] (afsoonalidadishamsa@sce.carleton.ca) is a PhD Candidate in the Carleton-NTN (Non-Terrestrial Networks) Lab in the Systems and Computer Engineering Department at Carleton University, Canada. Her research focuses on the non-terrestrial networks and the innovative integration of High Altitude Platform Stations (HAPS) with terrestrial wireless networks.
\end{IEEEbiographynophoto}
 \vspace{-33pt}
\begin{IEEEbiographynophoto}{Animesh Yadav}
    [SM] (yadava@ohio.edu) is an assistant professor in the School of EECS at Ohio University, Athens, OH, USA. He is the Senior Editor for IEEE Communications Letters and Associate Editor for Frontiers in Communications and Networks.
\end{IEEEbiographynophoto}
\vspace{-33pt}
\begin{IEEEbiographynophoto}{Yasser Gadallah}
    [SM] (ygadallah@aucegypt.edu) is a Professor at the Department of ECE at The American University in Cairo (AUC), Egypt. His research interests include IoT, wireless communication, and smart systems. He has extensive telecommunications industrial experience, having worked at several high-profile industry-leading companies in this area. 
\end{IEEEbiographynophoto}
\vspace{-33pt}
\begin{IEEEbiographynophoto}{Halim Yanikomeroglu} [F] (halim@sce.carleton.ca) is a Chancellor’s Professor in the Department of Systems and Computer Engineering at Carleton University, Canada, and he is the Director of Carleton-NTN (Non-Terrestrial Networks) Lab. His group’s focus is the wireless access architecture for the 2030s and 2040s, and non-terrestrial networks. He is a Fellow of EIC (Engineering Institute of Canada), CAE (Canadian Academy of Engineering), and AAIA (Asia-Pacific Artificial Intelligence Association).

\end{IEEEbiographynophoto}
\vfill

\end{document}